\documentclass{PoS}
\usepackage{subfig,amsmath}

\title{Systematic Errors of the MCRG Method}

\ShortTitle{Systematic Errors of the MCRG Method}

\author{Simon Catterall\\
        Syracuse University, USA\\
        E-mail: \email{smc@physics.syr.edu}}

\author{Luigi Del Debbio\\
        Edinburgh University, UK\\
        E-mail: \email{luigi.del.debbio@ed.ac.uk}}

\author{Joel Giedt\\
        Rensselaer Polytechnic Institute, USA\\
        E-mail: \email{giedtj@rpi.edu}}

\author{\speaker{Liam Keegan}\\
       Edinburgh University, UK\\
       E-mail: \email{liam.keegan@ed.ac.uk}}

\abstract{
We present a Monte Carlo Renormalisation Group (MCRG) study of the SU(2) gauge theory with two Dirac fermions in the adjoint representation. Using the two--lattice matching technique we measure the running of the coupling and the anomalous mass dimension. We find slow running of the coupling, compatible with an infrared fixed point. Assuming this running is negligible we find a vanishing anomalous dimension, $\gamma=-0.03(13)$, however without this assumption our uncertainty in the running of the coupling leads to a much larger range of allowed values, $-0.6 \lesssim \gamma \lesssim 0.6$. We discuss the systematic errors affecting the current analysis and possible improvements.
}

\FullConference{XXIX International Symposium on Lattice Field Theory \\
                July 10 -- 16 2011\\
                Squaw Valley, Lake Tahoe, California}

\begin{document}

\section{Introduction}

Technicolor theories with fermions in higher representations of the gauge group can potentially provide a dynamical electroweak symmetry breaking mechanism without conflicting with electroweak precision data. Minimal Walking Technicolor is an example of such a theory, a SU(2) gauge theory with two Dirac fermions in the adjoint representation~\cite{Sannino:2004qp,Luty:2004ye}. It is expected from perturbation theory to be in or near to the conformal window, and lattice simulations have indeed found that the gauge coupling runs very slowly~\cite{Bursa:2009we,DelDebbio:2009fd,Hietanen:2009az,DeGrand:2011qd,Giedt:2011kz}. In order to be phenomenologically viable the theory must have a large anomalous mass dimension $(\gamma\sim1)$~\cite{Holdom:1981rm,Yamawaki:1985zg,Appelquist:1986an}. A conjectured all--order beta function~\cite{Pica:2010mt} predicts $\gamma=11/24\simeq0.458$, which is consistent with perturbative results in the $\overline{\mathrm{MS}}$--scheme up to four loops~\cite{Ryttov:2010iz}, and recent non--perturbative lattice results~\cite{Bursa:2009we,DeGrand:2011qd,DelDebbio:2010hx}.

In this work we measure the running of the coupling and the anomalous mass dimension using the MCRG two--lattice matching method, a technique which has recently been used to investigate theories with many flavours of fermions in the fundamental representation~\cite{Hasenfratz:2009ea,Hasenfratz:2010fi,Hasenfratz:2011xn}.

\section{Two--Lattice Matching Method}

We consider the evolution of all possible couplings of the system under Wilson Renormalisation Group (RG) block transformations of scale $s$, where with each blocking step ultraviolet (UV) fluctuations are integrated out, and irrelevant couplings flow towards their fixed point (FP) values. After a few such steps the only remaining flow is in the relevant couplings along the renormalised trajectory (RT). If we can identify two sets of couplings whose actions are matched after $n$ and $(n-1)$ RG steps then their lattice correlation lengths must differ by a factor $s$. Since the physical correlation length should not be changed by the RG transform, this means that their lattice spacings, and hence inverse cutoffs $\Lambda^{-1} \sim a$ must also differ by a factor $s$. To identify such a pair of couplings, we need to show that after $n$ and $(n-1)$ RG steps respectively their actions are identical, which can be done by performing RG blocking on the gauge configurations and showing that the expectation values of all observables on these blocked gauge configurations agree.

We use three $s=2$ RG blocking transforms, labelled ORIG, HYP and HYP2~\cite{Hasenfratz:2009ea}. Each RG transform contains a free parameter $\alpha$, and changing $\alpha$ changes the location of the FP, and how quickly the RT is approached in a given number of steps. Ideally it should be chosen such that
\begin{itemize}
 \item All operators predict the same matching coupling between $(n,n-1)$ pairs for a given blocking step $n$. (Deviations are a measure of the systematic error from not being at exactly the same point along the RT.)
 \item Consecutive blocking steps predict the same matching coupling, i.e. the coupling for which $(n,n-1)$ pairs agree should be the same for all $n$. (Deviations show that the RT is still being approached in the irrelevant directions\footnote{If there existed other relevant couplings that had not been precisely tuned to their critical values, these deviations could also indicate an unwanted flow in these relevant directions}.)
\end{itemize}

\section{Pure Gauge Results}

\begin{figure}[ht]
  \centering
\subfloat[Plaquette Matching]{\label{fig:PURE_plaq_a}\includegraphics[angle=270,width=7.5cm]{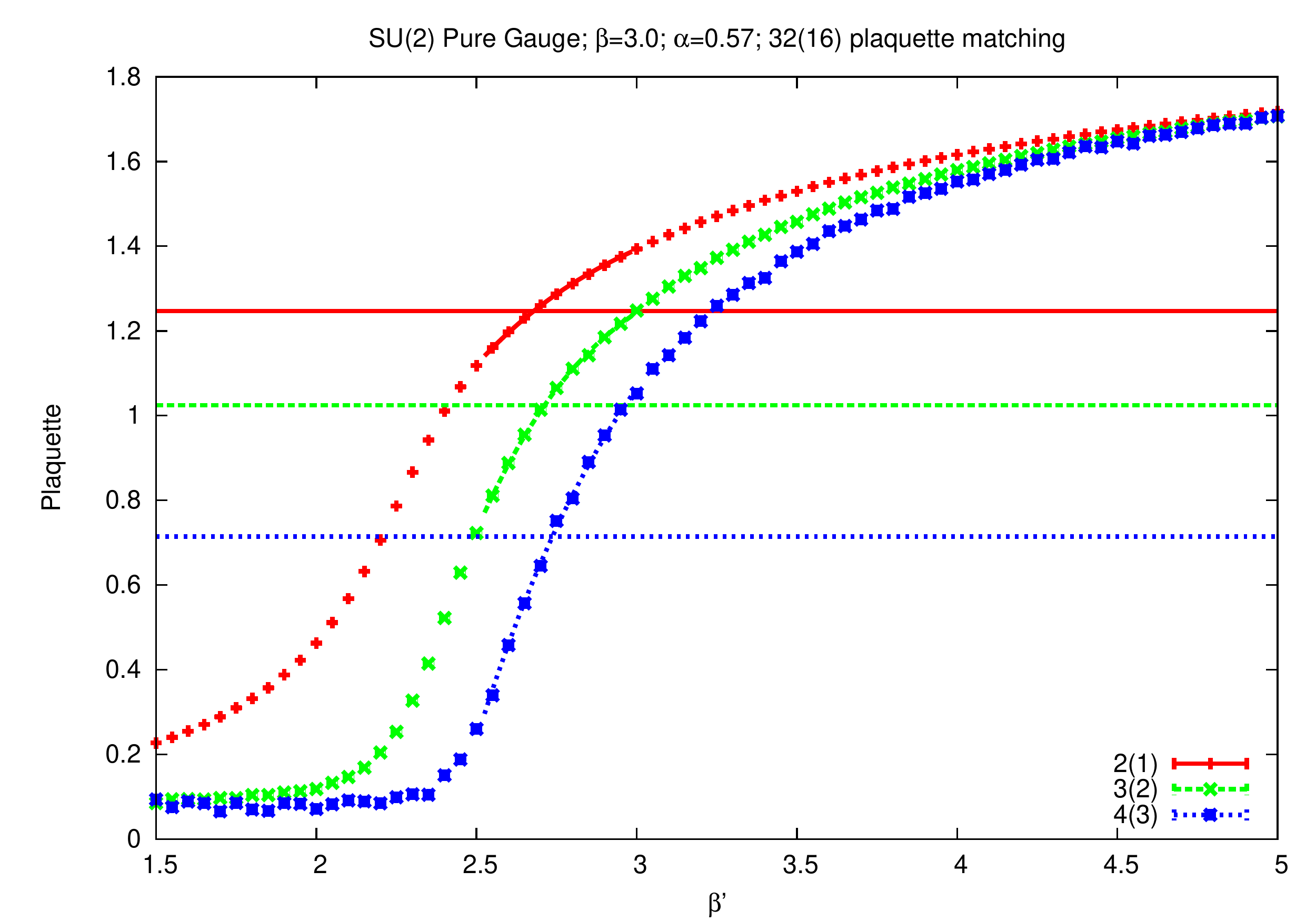}}
\subfloat[$\alpha$--Optimisation]{\label{fig:PURE_plaq_b}\includegraphics[angle=270,width=7.5cm]{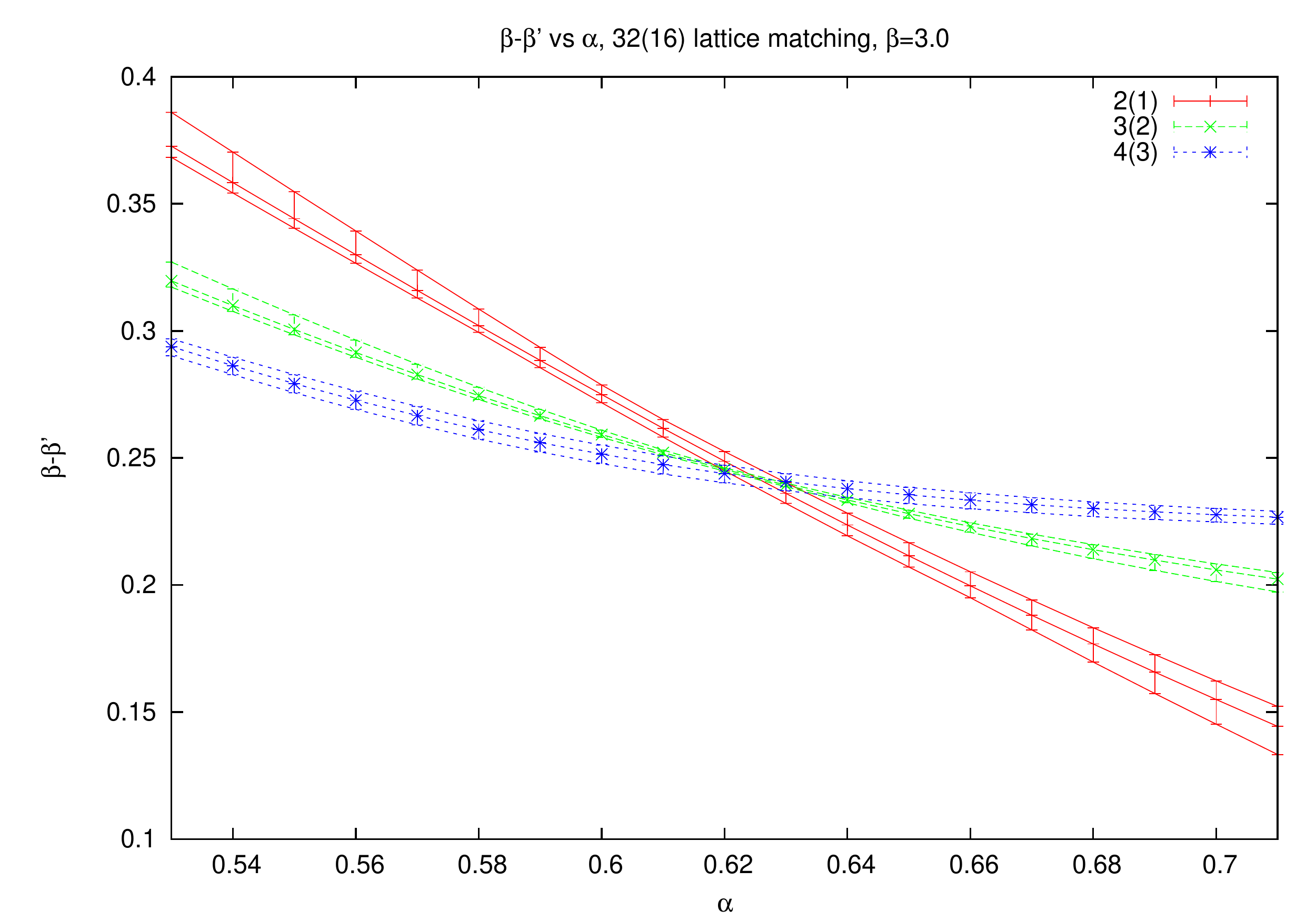}}
  \caption{An example of the matching of the plaquette in $\beta$ for the pure gauge case using ORIG blocking on 32(16) lattices, and the $\alpha$-optimisation including all seven matching observables.}
  \label{fig:PURE_plaq}
\end{figure}

As an initial test of the method, matching in $\beta$ between 32(16) and 16(8) lattices was performed using $\sim2000$ SU(2) pure gauge configurations for each $\beta$. We matched in the plaquette, the three six--link loops, and three 8--link loops. An example of the matching of the plaquette is shown in Fig.~\ref{fig:PURE_plaq_a}. The red, green and blue horizontal lines show the average plaquette on the $32^4$ lattice after 2, 3 and 4 ORIG blocking steps respectively, at $\beta=3.0, \alpha=0.57$. The interpolated red, green and blue points show the average plaquette on the $16^4$ lattice after 1, 2 and 3 blocking steps respectively, as a function of $\beta'$, also at $\alpha=0.57$. The value of $\beta'$ where the two red lines intersect gives the matching coupling for $n=2$, similarly the green and blue lines give the matching coupling for $n=3$ and $4$.

\begin{figure}[ht]
  \centering
\subfloat[Bare step scaling $s_b$ for the pure gauge theory]{\label{fig:PURE_sb}\includegraphics[angle=270,width=6.95cm]{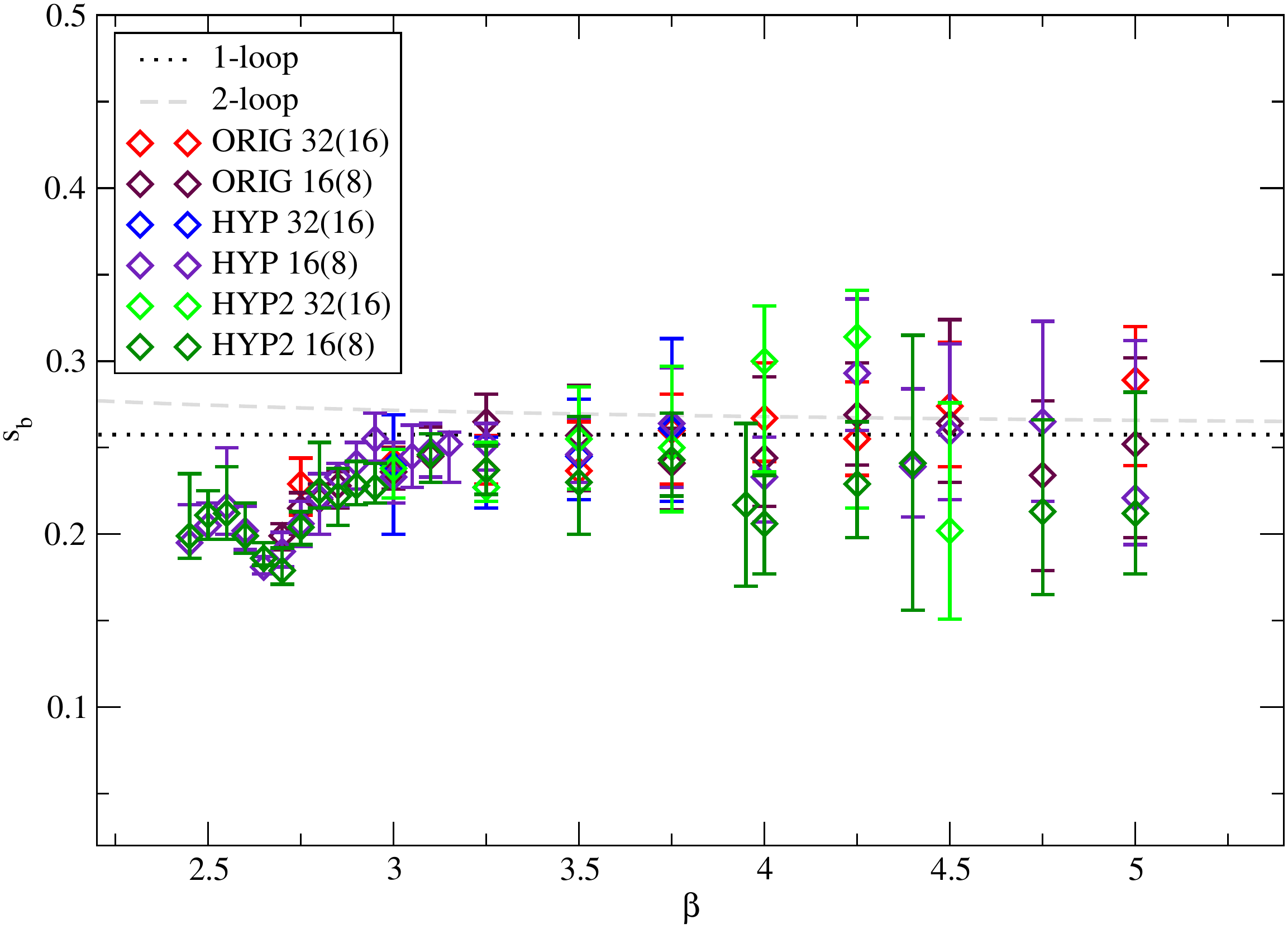}}
\subfloat[Bare step scaling $s_b$ for the full MWT theory]{\label{fig:MASSLESS_sb}\includegraphics[angle=270,width=7.2cm]{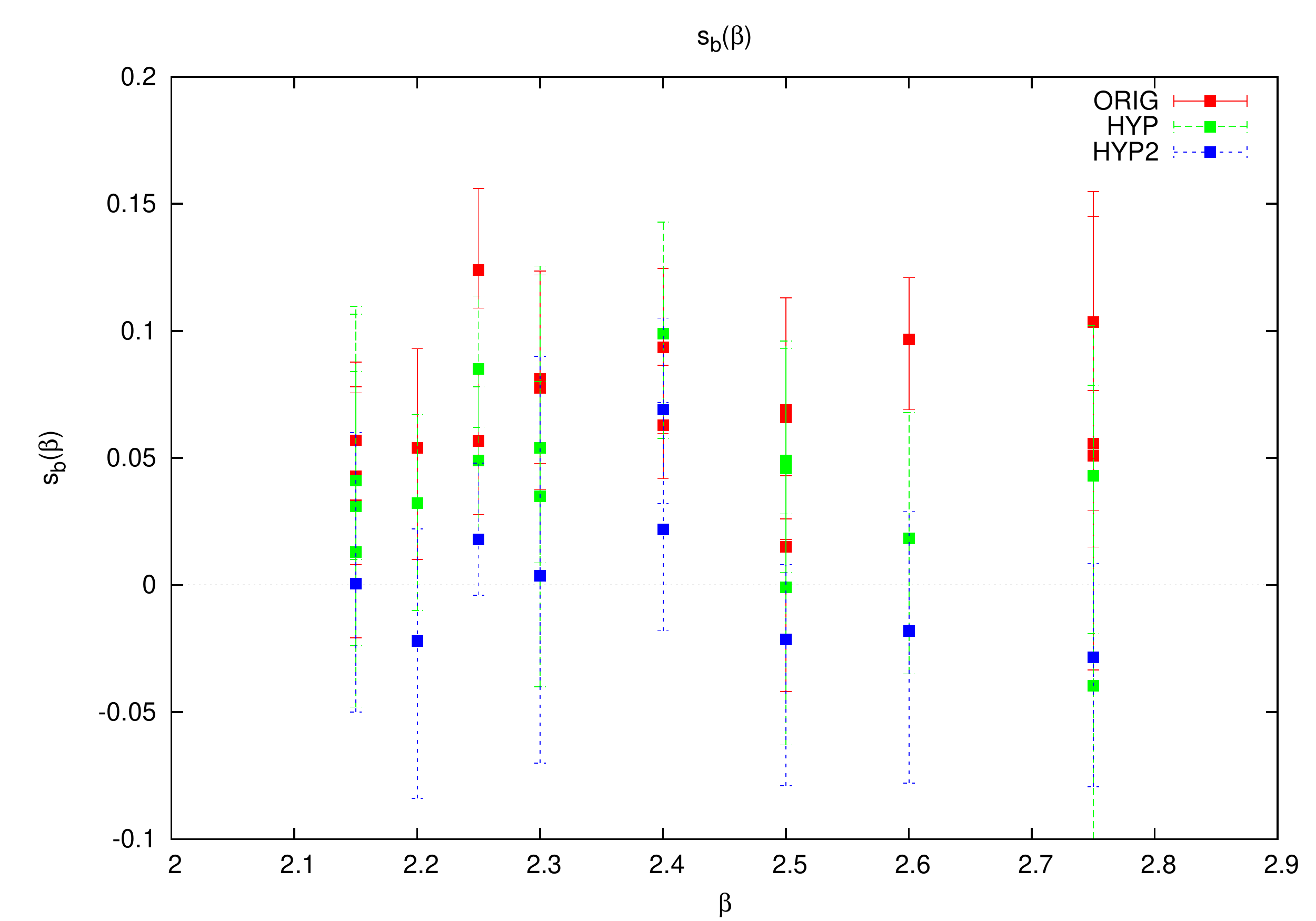}}
  \caption{The bare step scaling $s_b$ for the pure gauge theory (consistent with perturbation theory), and the full MWT theory (consistent with a FP), using ORIG, HYP and HYP2 blocking.}
  \label{fig:sb}
\end{figure}

This matching is repeated for each observable, and the spread of predicted matchings for each $n$ gives a systematic error on the central matching value. The whole procedure is then repeated for various values of $\alpha$, as shown in Fig.~\ref{fig:PURE_plaq_b}, to find an optimal value of $\alpha$ where subsequent RG steps predict the same matching value. The intersection of the last two blocking steps gives a central value for $s_{b}(\beta=3.0;s=2) \equiv \beta - \beta'$, while the range of couplings for which any of the blocking steps intersect within errors gives the uncertainty on this central value. Fig.~\ref{fig:PURE_sb} shows the resulting step scaling of the bare coupling $s_b$, determined using ORIG, HYP and HYP2 blocking on both 32(16) and 16(8) lattices, along with the 1--loop and 2--loop perturbative predictions.

\section{Discrete Beta Function Results}
\label{sec:coupling}

Having confirmed that the method works for the pure gauge case, we now turn to the full Minimal Walking Technicolor theory of two Dirac fermions in the adjoint representation of SU(2). There are two couplings of interest in this theory, the gauge coupling and the mass. At an IRFP the gauge coupling is expected to be irrelevant, leaving the mass as the only relevant operator. In order to measure $s_b$ we first need to tune to the mass to its critical value, so that we are following the running of the gauge coupling and not the mass, then the matching can be done in exactly the same way as for the pure gauge case. A fixed point would be indicated by a change of sign in the quantity $s_b$ as the bare coupling is varied from weak to strong coupling.

We use the HiRep~\cite{DelDebbio:2008zf} implementation of the Wilson plaquette gauge action with adjoint Wilson fermions and an RHMC algorithm with two pseudofermions. We then generated $\sim3000$ configurations on $16^4$ and $8^4$ lattices for a range of $\beta$ values, each run at the critical bare mass. This allows two matching steps, after $2(1)$ and $3(2)$ steps on the $16^4(8^4)$ lattices respectively.

The resulting measurement of $s_b(\beta)$ is shown in Fig.~\ref{fig:MASSLESS_sb}. This includes both the massless and some small mass ($am\sim0.005$) runs; within errors $s_b$ shows no mass dependence for these small masses. The ORIG matching values of $s_b$ are clearly positive throughout, the HYP values are lower, and the HYP2 values are consistent with zero within error bars. There is no clear cross--over from positive to negative values of $s_b$ for any of the blocking transforms, so while the data are consistent with a fixed point, they are not sufficiently precise to distinguish slow running from a fixed point.

\section{Anomalous Dimension Results}

Since the mass is the only revelant coupling at the IRFP, we should in principle be able to match in the mass at arbitrary gauge couplings, given sufficient RG steps for the gauge coupling to flow to its FP value, and a mass small enough that despite the resulting flow in the mass the system remains close to the IRFP. In practice we only have a small number of RG steps, and because the beta function is small the coupling only flows slowly towards its FP value, so we set $\beta'=\beta$ assuming the flow in the coupling due to one RG step is negligible. We match observables as for the pure gauge case, but instead of matching in $\beta$ we match pairs of bare masses $(am_0,a'm_0')$.

We generated $\sim3000$ configurations on $16^4$ and $8^4$ lattices, for a range of bare masses at each $\beta$. This allows two matching steps, after $2(1)$ and $3(2)$ steps on the $16^4(8^4)$ lattices respectively.

\begin{figure}[ht]
  \centering
\subfloat[Mass matching assuming $\beta=\beta'$]{\label{fig:MASS_HYPall}\includegraphics[angle=270,width=7.5cm]{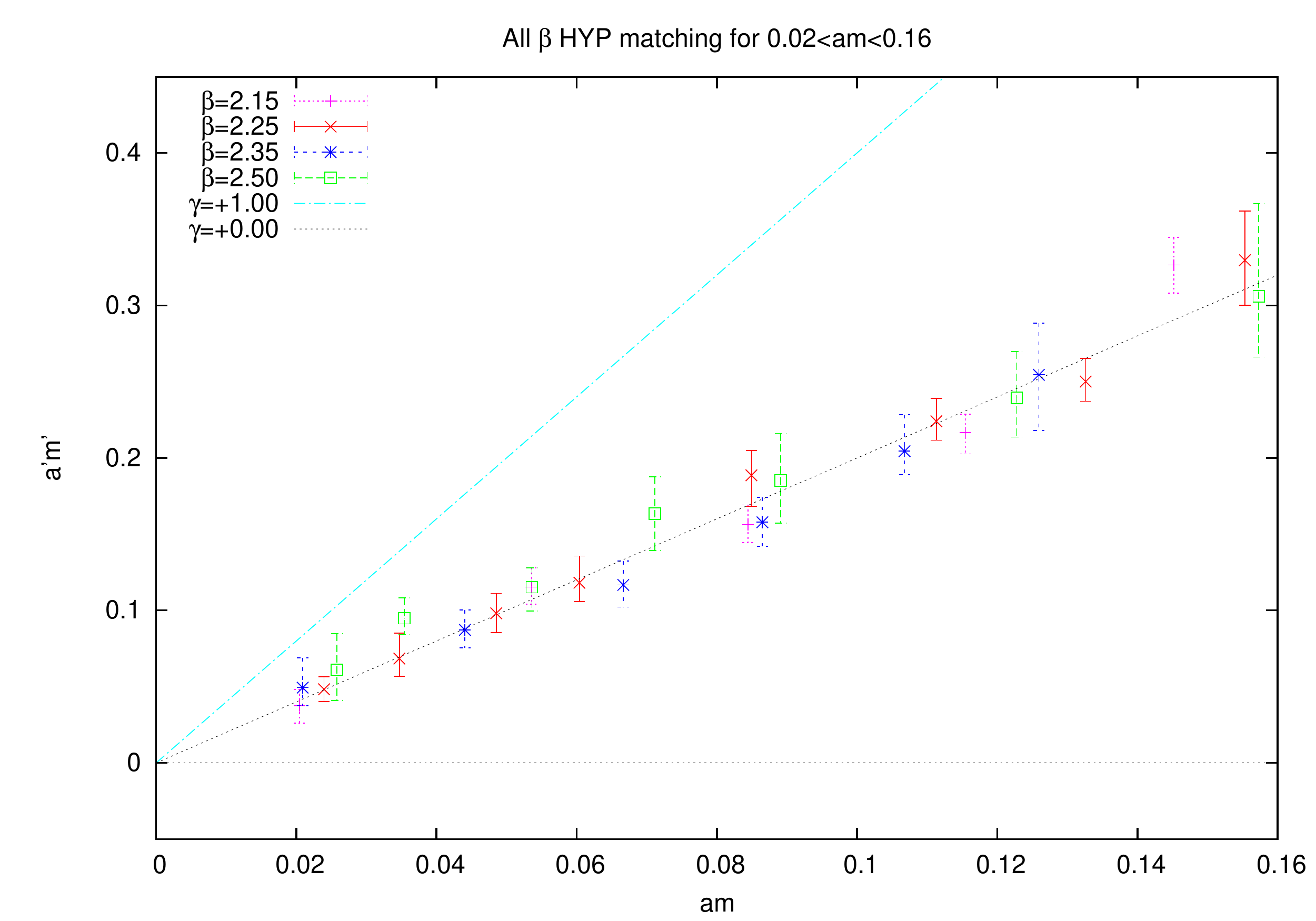}}
\subfloat[Mass matching at $\beta=2.25$ for a range of $\beta'$. ]{\label{fig:MASS_beta}\includegraphics[angle=270,width=7.5cm]{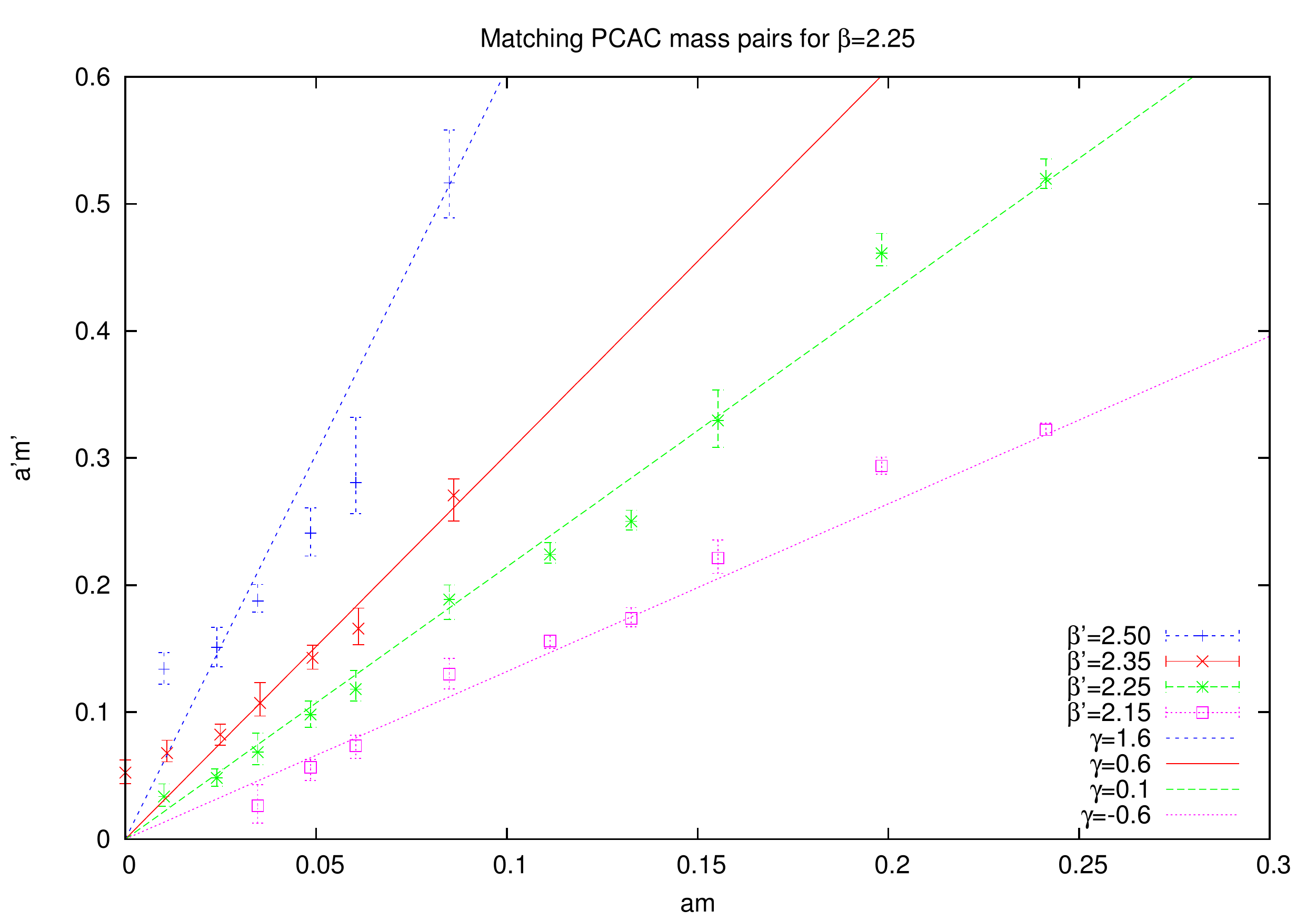}}
  \caption{Matching masses in the range $0.02<am<0.16$, with the assumption that $\beta'=\beta$, gives a vanishing anomalous mass dimension $\gamma=-0.03(13)$. However allowing $\beta'$ to vary increases the allowed range of values to $-0.6 \lesssim \gamma \lesssim 0.6$.}
  \label{fig:MASS}
\end{figure}

Because the bare mass is additively renormalised we convert the bare masses to PCAC masses. We measure the PCAC mass, $am$, as a function of bare mass, $am_0$, for each $\beta$ on the $16^4$ lattices. We then use this to convert the bare masses on both $8^4$ and $16^4$ lattices to PCAC masses, as the measured PCAC masses on the $8^4$ lattices suffer from finite volume effects. Our previous result~\cite{Catterall:2010du} for the anomalous mass dimension used PCAC masses measured on the $8^4$ lattices and hence contained a large finite volume effect, which has been removed in the present work.
The anomalous mass dimension appears in the RG equation for the mass,
\begin{equation}
\frac{d (am)}{d\ln|\mu|} = -y_m am = -(1+\gamma)am.
\end{equation}
At an IRFP the anomalous mass dimension is a constant, so this can be integrated to give
\begin{equation}
\label{eq:gamma}
\frac{a'm'}{am} = 2^{\gamma+1}
\end{equation}
for a pair of matching masses $(am,a'm')$, from which a value for $\gamma$ can be extracted. We used four values of $\beta$, $\beta=2.15,2.25,2.35,2.50$, and the matching PCAC mass pairs using the HYP blocking tranform are shown in Fig.~\ref{fig:MASS_HYPall}. We also repeated the matching using ORIG and HYP2 blocking, and the agreement between the three blocking methods is very good.

Different $\beta$ values seem to predict consistent values for the anomalous mass dimension, as shown in Fig.~\ref{fig:MASS_HYPall}, which uses all the $\beta$ values and masses in the range $0.02<am<0.16$. A combined fit to these data gives $\gamma=-0.03(13)$, however significant systematic errors exist that have not been accounted for in this result, in particular the assumption that we can set $\beta'=\beta$.

\section{Matching Observables}
\label{sec:systematics}
In principle, for a given $(\beta,am)$ there should be a unique matching set of couplings $(\beta',a'm')$, for which all blocked observables agree after $n$ and $(n-1)$ blocking steps respectively. In practice however, all of our observables are small Wilson loops, and as such are strongly correlated and have a very similar dependence on $\beta'$ and $a'm'$. This means that we can in fact find a ``matching'' $a'm'$ for a range of values of $\beta'$. As an example the matching mass pairs for $\beta=2.25$ and various values of $\beta'$ are shown in Fig.~\ref{fig:MASS_beta}. While we find that $s_b=\beta-\beta'$ is compatible with zero, corresponding to setting $\beta=\beta'$, the error bars are relatively large, $-0.08\lesssim \beta-\beta' \lesssim 0.16$. From Fig.~\ref{fig:MASS_beta} we see that for $\beta=2.25$ this region is approximately bounded by $\beta'=2.15$ and $\beta'=2.35$, and encloses a large range of values for the anomalous mass dimension, $-0.6 \lesssim \gamma \lesssim 0.6$. This range is representative of the errors in the anomalous mass dimension due to the uncertainty in the correct value of $\beta'$, and is the dominant source of systematic uncertainty in our results.

To resolve this issue, more observables which are ``orthogonal'' to the Wilson loops, for example meson correlators or other fermionic observables such as the PCAC mass, would need to be included in the matching to determine which of these blocked configurations are actually matched, giving a unique matching set of couplings.

\section{Finite Volume}
The calculation of $\gamma$ requires us to compute the flow in the mass parameter for a conformal field theory (CFT) in the presence of a small mass deformation. Of course to be a true CFT the theory must be considered in infinite volume but unfortunately when we do simulations we have lattices of finite size, and so both the mass parameter and the lattice size then determine any correlation length~\cite{DelDebbio:2010ze}. To extract the correct physics it is important to ensure that the correlation lengths we are measuring and matching are not being strongly influenced by the finite box size, $m \gg 1/L$, however if we use too large a mass we will move the
system a long way from any IRFP and the simple MCRG techniques we are using will not apply. For our L=16 lattices to achieve for example $mL > 2$ we would need $am\gtrsim0.12$, which may be simply too large a mass to keep the system close to the fixed point.

\section{Conclusion}
We find a vanishing anomalous dimension and slow running of the coupling consistent with an IRFP, however our results suffer from considerable systematic errors. The most significant of these is the uncertainty in the location of the fixed point in the coupling, which in turn produces a large uncertainty in the value of the anomalous mass dimension. This is essentially the same issue that the Schr\"odinger Functional studies of this theory have suffered from.

Assuming negligible running in the coupling we find $\gamma = -0.03(13)$.
Using our errors in the
measurement of the discrete beta function to suggest a window for
the possible flow in $\beta$ produces a large uncertainty in the associated
anomalous dimension $-0.6 \lesssim \gamma \lesssim 0.6$. A small value of $\gamma$ is also
likely to result from finite volume effects; if the beta function
is small but non--zero then many
blocking steps and hence large starting volumes
will be needed to move the system away from the usual
perturbative UV fixed point.

Adding more matching observables, in particular fermionic ones such as meson correlators, should give a broader set of observables which will enable us to determine unambiguously matched actions. Using $32^4$ and $16^4$ lattices matching could be done after 1/2 and 2/3 blocking steps - the smallest blocked lattice being of size $4^4$, so that for example the meson correlator at each timeslice could be used as an observable. Going to $32^4$ lattices will also allow us to go to smaller masses and thus be closer to the IRFP, as well as allowing us to match our current set of blocked observables after three blocking steps instead of two, which is a much more stringent matching condition that should help to constrain $\beta'$. We will also be able to check for finite volume effects using the analysis of Ref.~\cite{Hasenfratz:2011xn}, and it may be possible to get a determination of the anomalous mass dimension using the stability matrix MCRG method. Other potential improvements include using an improved action to reduce ${\mathcal O}(a)$ effects, and adding an adjoint plaquette term to the action to move the system away from a non--physical ultraviolet fixed point due to lattice artefacts, and allow us to go to stronger coupling~\cite{Hasenfratz:2011xn}.

However, even taking into account our currently large systematic errors, we find a small anomalous mass dimension which is clearly less than 1, and hence that, at least in its simplest form~\cite{Fukano:2010yv}, the SU(2) gauge theory with two Dirac fermions in the adjoint representation is not a viable Walking Technicolor candidate.

\end{document}